# Six Hypotheses for Accelerating the Lunar Economy

Michael Nayak[1]

1: Strategic Technologies Office, Defense Advanced Research Projects Agency (DARPA)

**Abstract:**

Based on technical work and development conducted under the LunA-10 study, I have identified six hypotheses where, if revolutionary improvements in technology can be made, I assess that a direct acceleration to the fielding of a lunar economy is likely to occur. In this short paper, I explain these six hypotheses, and recommend that these topics be focused on for technical development in the near-future by government and commercial stakeholders. These areas are:

(1) Centralized thermal rejection and generation as a service,
(2) Widespread orbital lunar prospecting and surveying,
(3) Creating large silicon wafers for microsystems on the Moon,
(4) Biomanufacturing to accelerate lunar construction,
(5) New concepts to increase refinement rates in low gravity,
(6) New concepts for Lunar Position, Navigation and Timing.

## 1. Introduction

DARPA supports a future model where the National Aeronautics and Space Administration (NASA), international governments, and commercial industry can rapidly scale up lunar exploration and commerce, enabled and supported by the deployment of an efficiently combined, integrated lunar infrastructure framework. An integrated framework would upend the current technical paradigm, whereby each lunar lander or activity must organically support all required resources for survival such as power, communications, and data storage.

To support NASA civil space and US commercial space goals, in November 2023, DARPA launched the 10-Year Lunar Architecture (LunA-10) study[1]. LunA-10 studied the business and technical case for non-terrestrial technology concepts designed to move away from individual scientific efforts within isolated, self-sufficient systems and toward a series of shareable, scalable, resource-driven systems that can operate jointly. The program focused on creating monetizable services for future lunar users in a mass-efficient manner, while complementing existing NASA and international partner (Artemis Accords[2]) lunar investments.

Based on technical work and development conducted under the LunA-10 study, I have identified six hypotheses where, if revolutionary improvements in technology can be made, a direct acceleration to the fielding of a lunar economy may be likely to occur.

## 2. Topic #1. Centralized Thermal Rejection and Generation as a Service.

---

[1] https://sam.gov/opp/9ef8be0d69df4f7d9dd8077e411a852a/view
[2] The *Artemis Accords* describes a shared vision for principles, grounded in the Outer Space Treaty, to create a safe and transparent environment which facilitates exploration, science, and commercial activities for all of humanity to enjoy. Details are available at https://www.nasa.gov/specials/artemis-accords/index.html.

A commercial, self-sustaining lunar economy will include services that operate through both lunar day and lunar night, at both the equator and Poles. This will require each lunar system to include mechanisms for both thermal rejection (lunar day) and thermal generation (lunar night survival). These extremes drive the necessity for large thermal systems that include heavy components such as radiators, heat pipes, pumps, heat exchangers, coolant fluids, thermal controllers, etc.

This results in operating lunar systems where a significant mass fraction is thermal rejection or generation. For example, the mass fraction of a thermal system, as a percentage of total system mass, can be as high as 60% for an optical-wavelength power beaming system, 45% for an In-situ Resource Utilization (ISRU) plant, and 25% for surface fission plants. Proton exchange membrane fuel cells (PEMFC) output up to 45% heat[3]. Thermoelectric devices such as Radioisotope thermoelectric generators (RTG) have electrical conversion efficiencies between 5-20%; ~300 W of generated electrical power can create continuous heat generation levels as high as 4-5 kWt[4]. System hardware needed to reject this waste heat reduces mass efficiency for the overall system that use RTGs or PEMFCs for power. Fundamentally, lunar expansion to infrastructure-level services is limited by this mass paradigm.

Instead, if thermal rejection/generation can be offloaded from individual systems and centralized in a thermal hub, thereby eliminating upto 60% of an individual system's mass, I hypothesize that this would represent both a large financial advantage[5] for commercial companies, and a drastic scaleup in mission capability (power, robotics, construction, etc.) for the same mass delivered to the Moon. This future thermal hub is envisioned to work analogously to a heating, ventilation and air conditioning (HVAC) system in a commercial building. Each floor in a commercial building might have its own heating or cooling requirements, but all are managed by a single building-wide HVAC unit. The centralized thermal rejection/generation hub would establish connections to an arbitrary number of nearby users – during the lunar day, the hub would push cooling fluid and use its heat pumps to draw heat away from users, and during the lunar night, release stored or generated heat to those same users for survival needs.

Consolidating thermal components of users in a geographical area into one hub could eliminate the need for multiple sets of thermal hardware. For example, larger heat pumps deliver more power per unit mass; one large heat pump weighs significantly less than multiple smaller heat pumps, reducing overall mass to the lunar surface and favorably scaling to growing numbers of users. Consolidation also permits the hub's thermal design to be sized to manage *average* thermal requirements across *all* users, as opposed to individual system-level *peak* thermal requirements for a single user, which could result in greater mass savings.

This mass savings can then become a commercial utility service. In this new paradigm, users would bring only the minimal thermal equipment to the Moon, plug into a thermal hub, and pay for heat generated/rejected on a dollars-per-kilowatt basis – analogous to power utilities on Earth and a foundational enabler to the lunar economy.

This hub would be directly in line with the LunA-10 goals of "moving away from individual efforts within isolated, self-sufficient systems and toward a series of shareable, scalable, resource-driven systems that can operate jointly".

Some important technical issues that remain to be understood, and are the purview of future research I advocate for, are as follows:
- Specific highly-innovative method(s) by which drastically increased thermal rejection per unit mass

---

[3] DOI 10.1016/j.pecs.2021.100966.
[4] kWt: Kilowatts thermal. kWe: Kilowatts electrical.
[5] Estimates for cost to the lunar surface vary between $1M-$1.5M USD per kilogram.

- may be possible, by radiation, active cooling, conduction to the lunar surface, or other.
- Specific highly-innovative method(s) by which drastically increased thermal storage per unit mass may be possible for heat gathered from users.
- The overall possible mass savings by thermal aggregation (heat rejection and heat generation), and how this scales from tens to hundreds of kWt of heat rejected/generated.
- The possibilities of the thermal hub extracting useful electrical energy from excess heat extracted from users, and how this impacts overall hub mass efficiency.
- The application of recent material science developments (in thermal control) and how they might integrate into a system-level thermal hub, in such a way that an order-of-magnitude decrease in mass delivered to the Moon might be realized.

### 3. Topic #2. Widespread Lunar Orbital Prospecting

Lunar resources are pivotal in enabling a sustained human presence on the Moon and for developing a vibrant cislunar economy[6]. There remains uncertainty about the true commercial value of resources available on the Moon, and the resulting viability of the lunar economy. Current knowledge points to water and oxygen extractable from cold traps close to the lunar poles, but other resources, to include those that may be present beneath the top meter of lunar regolith, are unknown. Data of sufficient fidelity are not available to understand the distribution of resources across the Moon at the near sub-surface level (defined as the top 3-10 meters of the Moon). Once those areas are identified, then reserves of economically recoverable material could be evaluated for mining and processing operations, and galvanize the economics behind a future lunar economy.

To obtain informing datasets, methods by which to conduct a single-orbiter campaign of very low altitude (defined as below 15 km from the surface of the Moon) for resource evaluation, prospecting and lunar surveying are necessary, in order to quantify the widespread nature, abundance, accessibility and extractability of lunar resources[7]. A single orbiter architecture is considered for practicality and cost reasons. To accomplish this mission with one orbiter requires a singular ability to maneuver by using very low amounts of propellant or none at all, while also hosting spectrometer or other payloads capable of conducting resource prospecting.

Such an approach with alternate payloads can further be leveraged to help with lunar infrastructure surveying, and has the potential to build 3D photogrammetry models of the lunar surface that do not exist today. Building models of regions planned for active exploration supports virtual mission planning activities and facilitation of efficient, safe operations on the lunar surface.

Some important technical issues that remain to be understood, and are the purview of future research I advocate for, are as follows:
- Specific highly-innovative method(s) by which drastically increased low-altitude orbital station-keeping and maneuvering, for coverage anywhere on the lunar surface, may be possible.
- Sensing modalities best suited to support the prospecting and surveying activity, and the design of a total system that is set up to rapidly capture, disseminate and/or process large data sets.
- Relevant data and accuracy necessary to justify a next step in prospecting activity.
- For a photogrammetry mission, sensing modalities and resolution required to support surveying on the lunar surface.

---

[6] See, for example:
https://www.nasa.gov/sites/default/files/atoms/files/a_sustained_lunar_presence_nspc_report4220final.pdf
https://ntrs.nasa.gov/api/citations/20220005087/downloads/HEOMD-006_2022-03-25_FINAL%20033022.pdf
[7] DOI 10.1016/j.actaastro.2023.11.017

### 4. Topic #3. Creating Large Silicon wafers for microsystems on the Moon.

Due to different hydrostatic pressure, thermal convection and buoyancy conditions in space than on Earth, crystals grown in microgravity (0 g) have been shown to be up to 60% larger, up to 75% higher quality and up to 80% more uniform than those grown on Earth; a literature review of crystals grown in microgravity showed an improvement across 90% of crystals analyzed in size, structure, uniformity, resolution limit and/or mosaicity . This can be leveraged via in-space manufacturing to create a growth system for silicon wafers, and ultimately commercial-scale wide bandgap and ultra-wide bandgap semiconductors, of unprecedented size and quality. For example, gas flow for sweeping oxides or infrastructure to support the weight of liquid melt are no longer required; the lack of hydrostatic pressure creates larger crystals, with collapse pushed out to larger maximum diameters; the lack of thermal convection in vacuum negates inhomogeneous composition defects.

However, manufacturing in zero-gravity can necessitate highly complex systems. Some gravity, although not as much as 1g, is beneficial. The Moon (1/6g) may present an advantageous compromise between minimally-modified, gravity-fed processes that share commonality with Earth-based methods, while being able to leverage the benefits of <1g manufacturing. In this new paradigm, where it is feasible to manufacture large (defined as >400 mm) Silicon wafers on the surface of the Moon, a post-Moore's law future becomes possible, while addressing critical needs for Silicon-based microsystems that form the bedrock of modern Earth-based technological society.

Some important technical issues that remain to be understood, and are the purview of future research I advocate for, are as follows:

- Specific highly-innovative method(s) by which lunar-based Silicon wafer manufacture may be possible.
- The microgravity crystal growth hardware, and first-order anticipated processes, in order to create lunar-manufactured Si wafers; these must address power/thermal supply assumptions.
- Silicon crystal growth occurs at 1425 deg Celsius, which is approximately the temperature at which multiple ISRU pilot plants intend to operate, e.g., for carbothermal reduction of oxygen from regolith. There may be potential to co-design oxygen/water/other ISRU plants for silicon wafer manufacture, either using 99% pure silicon sourced from Earth, purified lunar-sourced regolith[8], or waste heat from other sources of the lunar economy.
- The value case for lunar manufacture, by direct comparison of maximum size, purity, structure and/or uniformity that could be achieved using 1/6g (lunar gravity), compared to microgravity (0 g) and Earth gravity (1 g).
- Increase in growth speed and rate of defects for lunar-manufactured Si wafers over Earth-manufactured Si wafers.
- Efficiencies and production rates of scalable systems capable of lunar Si wafer or wide band-gap semiconductor manufacture.

### 5. Topic #4. Biomanufacturing to accelerate Lunar Construction.

Microbial biomanufacturing has the potential to provide integrated solutions for remote or austere locations[9]. In addition to mechanical/physical/chemical approaches, biotechnologies broadly (and microorganisms specifically) will help enable long-term activities. These activities may rely on In-Situ Resource Utilization (ISRU), manufacturing, and energy collection/storage, but the efficiency of biotechnology may be particularly disruptive when used with locally available resources and optimized toward closed loop systems. Loop-closure, which indicates the recycling and the reuse of resources toward the establishment of a circular economy, is key not only to minimize the costs of resupply of resources from Earth but also for ethical considerations associated with space waste generation and the preservation of environments[10].

---

[8] Lunar regolith is approximately 21% Silicon by average composition.
[9] DOI 10.1111/1751-7915.13927

Tailored in-space biomanufacturing has emerged as a promising approach by which to create bio-enabled structures from source material such as lunar regolith, produce industrial fuels and lubricants (given some Earth-sourced starting materials), or even biomine rare earth elements present in trace concentrations that do not justify industrial mining (see Section 6 of this paper). It may be feasible to create modular, autonomous biomanufacturing, toward closed-loop lunar systems.

Some important technical issues that remain to be understood, and are the purview of future research I advocate for, are as follows:

- The utilization of lunar regolith or other lunar sourced materials as a feedstock for bio-enabled structures.
    - LunA-10 explored the possibility of utilizing waste streams from commercial services for use by other services or recycling; there may be potential in using waste streams from other lunar-based services as a feedstock for biotechnology or biomanufacturing-based systems.
- Specific mobilizable feedstocks, such as regolith components or other feedstocks resulting from lunar infrastructure, that engineered microorganisms could utilize.
- Resource streams required for regolith bioremediation and biomining (including microbial hardware needs) informed by ISRU and loop closure, to at least a materials manufacturing output.
- Process flow diagrams that could feasibly result in buildable and testable biological systems and fermentation hardware by which to create bio-enabled structures, pharmaceuticals, fuels or lubricants.
- The viability of biomining rare earth elements, particularly for trace concentration elements where industrial mining may not be economically viable (see Section 6).
- Techno-Economic Analysis (TEA) of closed-loop bioregenerative systems with inputs such as recycling of media from fermenters, or other sources that might stem from a robotically-enabled lunar economy.
- Estimates for production rate of construction elements such as compacted regolith bricks, ingots or compact landing pads for heavy lift lunar landers as a function of biomanufacturing and bioregenerative systems.

## 6. Topic #5. New concepts to Increase Refinement Rates in Low Gravity.

Several elements present on the lunar surface and on asteroids may have large economic value, but trace concentration levels mean that extraction of these elements may not be economically viable. For example, the Procellarum KREEP terrane on the near-side of the Moon has 300 times more uranium and thorium than chondrites[11], but abundances are in the single digit parts per million (ppm) (Uranium (U): 2 ppm[12]; Thorium (Th): 7-10 ppm[13]).

The abundance of Platinum Group Metals (PGMs) on PGM-rich asteroids are between 100-250 ppm[14], which presents a more favorable extraction scenario. However, when the price per troy ounce for PGMs of

---

[10] DOI 10.1038/s41467-023-37070-2
[11] DOI 10.1029/1999JE001092
[12] DOI 10.1029/2010GL043061
[13] DOI 10.1088/1674-4527/19/6/76
[14] For example, DOI 10.1016/j.actaastro.2023.03.019

>0.99 fine is compared to the cost of launch and return to Earth, it is likely that at least 1 metric ton of return per mission would be needed for economic viability. At even a 250 ppm abundance, ~4,000,000 kg of raw material would need to be refined at the source. This is a significant amount of mass to move in micro- to lunar-gravity, but will also take significant time (order of 10 years) to process through a spacecraft-sized refinery payload.

Novel mining methods and/or system designs by which to drastically increase the throughput, beneficiation and refinement of regolith (whether lunar or asteroidal) for the extraction of elements that are in the <100 ppm range of abundance are therefore paradigm-shifting. Specifically, mining refinery input rates significantly in excess of 3 kg/minute/W (where W is watts of power provided to the system) would present a step change to the current state-of-the-art.

Some important technical issues that remain to be understood, and are the purview of future research I advocate for, are as follows:

- Specific highly-innovative method(s) and new insights by which drastically increased beneficiation or refinement rates, of at least 3 kg/min/W, may be possible in 0 to 1/6 g.
- Refinery mobility or feeding mechanisms required to ensure that mining operations can continue at a constant rate, when mining at the ~million kg scale. For example, apron or reciprocating feeders on the Earth ensure that material is transported to the refinement equipment and not vice versa, but in space the refinery and feeder may be part of the same spacecraft payload.
- Methods by which to deal with particle sizes ranging from fines to larger boulders, while maintaining a constant processing rate.
- Direct consideration to the unique vacuum/pressure/environmental factors for efficient in-space mining, and technical insights to create a quantifiable, direct improvement in refinement rates.

## 7. Topic #6. New concepts for Lunar Position, Navigation and Timing (PNT).

If one were to architect the Global Positioning System (GPS) today, would it be space-based? Or would it be a series of ground-based systems that provided corrections to users in the vicinity of a dense population center, similar to the Wide Area Augmentation System (WAAS)? Space allows for GPS signals to be broadcast to remote parts of the world that may not be located near dense population centers. With lunar commercial activity limited to a few small areas heavy in resources, such fidelity or even widespread nature is likely not necessary. It is therefore unlikely that an Earth-like solution to the Position, Navigation and Timing (PNT) problem is the best solution for the Moon.

Therefore, I believe it is time to inform radically new concepts for PNT specific to the lunar domain. In particular:

- Miniaturized, very-low Size Weight and Power (SWaP) solutions for generating, maintaining and sharing timing signals on the Moon, independent of signals from Earth.
- Very low SWaP concepts for maintaining lunar surface-based position and navigation to the ~10-meter accuracy level when not in line of sight of an orbital or ground-based asset.
- Ways by which to share, update and synchronize PNT signals between mobile users beyond line of sight from one another via innovative new methods, such as (for example) by using low-frequency radio waves for data transmission via ground waves[15].
- The true requirements for lunar PNT, given near-future commercial paradigms of operation.

---

[15] NASA/TM—2008-215463, and Bader et al (1977), "The propagation of radio waves along the lunar surface".

**Acknowledgements.**

This work was conducted at the Strategic Technology Office, DARPA. The LunA-10 DARPA and Government Integration Teams contributed significant thought leadership to this work. The views expressed are those of the author and do not reflect the official policy or position of the Defense Advanced Research Projects Agency, the Department of Defense or the US Government. The material in this paper is UNCLASSIFIED and approved for public release.